\documentclass[12pt]{article}
\usepackage{epsfig}

\parskip=\medskipamount % space between paragraphs (TeX)
plus.3em minus.5em \abovedisplayshortskip=.5em plus.2em minus.4em
\renewcommand{\thesection}{\arabic{section}}

\catcode`@=11

\@addtoreset{equation}{section} \@addtoreset{equation}{subsection}
\def\theequation{\ifnum\value{section}=0 \arabic{equation}\ignorespaces
\else \ifnum\value{section}=-1 A.\arabic{equation}\ignorespaces
\else \ifnum\value{subsection}=0
\thesection.\arabic{equation}\ignorespaces \else
\thesection.\arabic{subsection}.\arabic{equation}\ignorespaces
                             \fi
                        \fi
                   \fi}

{\catcode`\'=\active \def'{{}^\bgroup\prim@s}}

\catcode`@=12

\newcommand{\bq}{\begin{equation}}
\newcommand{\be}{\begin{equation}}
\newcommand{\fq}{\end{equation}}
\newcommand{\ee}{\end{equation}}
\newcommand{\bqr}{\begin{eqnarray}}
\newcommand{\beqs}{\begin{eqnarray}}
\newcommand{\fqr}{\end{eqnarray}}
\newcommand{\eeqs}{\end{eqnarray}}

\newcommand{\rf}[1]{(\ref{#1})}

\def\bop#1{\setbox0=\hbox{$#1M$}\mkern1.5mu
    \vbox{\hrule height0pt depth.04\ht0
    \hbox{\vrule width.04\ht0 height.9\ht0 \kern.9\ht0
    \vrule width.04\ht0}\hrule height.04\ht0}\mkern1.5mu}
\begin{document}
\thispagestyle{empty}

\begin{flushright}
\begin{tabular}{l}
% TEP- \\
hep-th/0504188 \\
\end{tabular}
\end{flushright}

\vskip .6in
\begin{center}

{\bf  Algebraic and Polytopic Formulation to Cohomology}

\vskip .6in

{\bf Gordon Chalmers}
\\[5mm]
% {\em address \\
%      address \\
% Los Angeles, CA } \\

{e-mail: gordon@quartz.shango.com}

\vskip .5in minus .2in

{\bf Abstract}

\end{center}

The polytopic definition introduced recently describing the topology of manifolds is 
used to formulate a generating function pertinent to its topological properties.  In 
particular, a polynomial in terms of one variable and a tori underlying this polynomial 
may be defined that generates an individual cohomological count.  This includes the 
de Rham complex for example, as well as various index theorems by definition such as 
homotopy.  The degree of the 
polynomials depends on the volume used to define the region parameterizing the 
manifolds; its potentially complex form and L-series is not presented in 
this work.  However, the polynomials and the relevant torii uniformize the topological 
properties in various dimensions; in various dimensions  
this is interesting in view of known topologies. 

\vfill\break

\noindent {\it Introduction} 

The classification of manifolds has occupied theorists, both physicists and 
mathematicians of many types, for a long time.  In general the construction of 
invariants such as the dimensions of de Rham complexes or the computation of 
indices of spectral operators relevant to homotopy or cobordism groups follows 
from a variety of techniques.  The dimension of the space (or manifold) is 
important for both the technique used to compute the invariants and the 
end result.  

Recently a polytopic construnction was introduced so that a set of points in 
a lattice consisting of $N^d$ entries may be parameterized by a single number $z$.  
This means that all $q_1$-dimensional manifolds embedded in $q_2$-dimensional 
space may be labeled by a single number.  Each number represents a manifold, and  
the construction permits possible uniformizations of the properties of manifolds 
in diverse dimensions, for example a quantity is computable via a function $P(z)$.  
The polytopes were given in \cite{ChalmersOne}; related work on the L-series is in 
\cite{ChalmersTwo}, and on knot invariants and polynomials in \cite{ChalmersThree}.  
Compact number expressions describing gauge amplitudes might also be relevant.

The invariants are introduced and defined in the subsequent section, and following 
this, the latticed manifolds are defined and numbered.

\vskip .2in 
\noindent {\it Invariants} 

Consider that a manifold in $d$ dimensions embedded in a lattice of $N_d$ points 
is labeled by a number from $0$ to $N$.  Then by definition a polynomial of 
degree $N$ 

\bqr 
P_I({\tilde z}) = \sum b_i {\tilde z}^i 
\label{polynomial}
\fqr 
generates an invariant $I$ for all manifolds placed in the space.  Example would 
be the individual $(p,q)$ forms in the Dolbeaux cohomology, homotopies, differential 
structures, etc...   Associated to the function \rf{polynomial} is a function,  

\bqr 
{\tilde z} = \sum c_i z^i \ ,
\fqr 
which is a map from the basis of the numbers labeling the manifolds to another 
set of numbers.  The latter set of numbers is potentially more convenient for 
the input into the polynomials in \rf{polynomial}.

The space of all the manifolds, each labeled by an integer, generates 
numbers by the manipulation, 

\bqr 
\pmatrix{ P_I(z_1) \cr P_I(z_2) \cr \ldots \cr P(z_{N_d}) }  =  
\pmatrix{ z_1 & z_1^2 & \ldots & z_1^N \cr 
  	  z_2 & z_2^2 & \ldots & z_2^N \cr 
	  \ldots & \ldots & \ldots & \ldots \cr 
	  z_{N_d} & z_{N_d}^2 & \ldots & z_{N_d}^N } 
\pmatrix{ b_1 \cr b_2 \cr \ldots \cr b_N }  \ . 
\label{polynomialexp} 
\fqr 
An explicit construction of the numbers $b_i$ via the left inverse show that for 
integer $P_I(z_i)$ the numbers may be rationalized into integers by multiplying them 
via the inverse of the matrix in \rf{polynomialexp}.  

The same set of numbers generates an L-series after the numbers $b_i$ are rationalized 
to integers and placed into correspondence with the primes via counting them in 
accordance with the integers $i$ via $p_i$, and with the $b_i$ numbers, 

\bqr 
\zeta(s,C) = \prod \left(1+a_p p^{-s} + p^{1-2s}\right)^{-1} \ ,   
\label{Lseries}
\fqr 
and $-p+b_p=a_p$.  The construction in terms of the L-series means that there is 
an elliptic curve defined for the lattice with $N$ points with $b_p$ solutions to 

\bqr 
y^2=x^3 + \alpha_1 x + \alpha_2 \quad {\rm mod}\quad p \ . 
\label{torus}
\fqr 
There is a series of curves for the series of $N$ points (and the dimensionality).  
Depending on the invariant the curves might have special properties.  

A separate modular form is defined by 

\bqr 
\sum e^{-P_I(z)} = \sum \Delta_I(w) e^{-w} \ ,  
\label{pseudoHecke}
\fqr 
which essentialy counts the distributions of the invariants of the embedded manifolds 
parameterized by the polynomial $P_I(z)$.  The function $\Delta_I(w)$ counts the 
number of $z$-solutions to $w=P_I(z)$.  

The direct computation of these invariants requires specifying the ordering of the 
points in the lattice defining the embedding space.  The exact ordering of the 
embedding of the lattice 
points and their relation to the number labeling the manifold may play an important 
role in specifying the forms of the polynomials and their associated modular forms,  
and tori.  In other words, an appropriate choice may simplify both the L-series and 
torus in \rf{Lseries} and \rf{torus} (and that of the 'pseudo'-modular form in 
\rf{pseudoHecke}).  

It is of interest to find the best parameterization of the embedded manifolds, via 
the integers $z$ and the mapping ${\tilde z}=f_I(z)$ relevant to the index $P_I(z)$.  
This is useful in various dimensions to both relate the $P_I$ quantities to more 
symmetric tori (symmetric via their L-series) in \rf{torus} and also to eachother 
in differing dimensions.  

The finite volume specification of the invariants $P_I$ are expected to have a 
$V\rightarrow\infty$ limit.

Although the exact form of the invariants are not given here, it would certainly 
be interesting to find their forms and any coefficient structure in the different 
dimensions, and at large volume.  The L-series formulation could be relevant in 
the classification of topologies.  Also, fractional dimensional manifolds can 
be characterized by continuation of the $P_I^{(d)}$ into non-integral dimensions; 
the continuation could define the properties in this case.   

\vskip .2in 
\noindent {\it Polytope construction}

In this section the polytopic, or rather simplicial complex, construction 
of the manifolds is given.  The topology of the manifolds are found essentially 
by filling in a lattice with a set of points.  A basic simplicial complex is 
defined by 'connecting the dots' in $R^d$.  In the polytopic construction here, 
space-filling 'membranes,' embedded in $R^d$, are used.  All embeddable manifolds 
are accessible via this construction (such as a Klein manifold in $d=4$).  

The polytopes (simplicial complexes) considered are constructed via
a set of integers that label the points and faces parameterizing the
surface.  The integers may be given a matrix representation that
permits a polynomial interpretation, and hence maps to knot(s)
invariant(s).

Take a series of numbers $a_1 a_2 \ldots a_n$ corresponding to the
digits of an integer $p$, with the base of the individual number being
$2^n$; this number $a_j$ could be written in base $10$ by the usual
digits.  In this way, upon reduction to base $2$ the digits of the
base reduced number spans a square with $n+1$ entries.  Each number
$a_j$ parameterizes a column with ones and zeros in it.  The lift of
the numbers could be taken to base $10$ with minor modifications,
by converting the base of $p$ to $10$ (with possible remainder issues
if the number does not 'fit' well).

The individual numbers $a_i$ decompose as $\sum a_i^m 2^m$ with
the components $a_i^m$ being $0$ or $1$.  Then map the individual
number to a point on the plane,

\bqr
{\vec r}_i^m  = a_i^m \times m {\hat e}_1 +
       a_i^m \times i {\hat e}_2 \ ,
\fqr
with the original number mapping to a set of points on the plane via
all of the entries in $a_1 a_2 \ldots a_m$.  In doing this, a collection
of points on the plane is spanned by the original number $p$, which
could be a base $10$ number.  The breakdown of the number to a set of
points in the plane is represented in figure 1.

A set of further integers $p_j=a_1^{(j)} a_2^{(j)} \ldots a_n^{(j)}$
are used to label a stack of coplanar lattices with the same procedure
to fill in the third dimension.  The spacial filling of the disconnected
polhedron is assembled through the stacking of the base reduced
integers.

Colored polytopes are introduced by taking the integers $p_j$ into
the numbers $a_j^{(k,m)}$ with base $N$ as opposed to base $2$.  The individual 
entries in the lattice are spanned by the vector,

\bqr
{\vec r}= {\vec r}_i^m  = a_i^{m,k} \times m {\hat e}_1 +
       a_i^{m,k} \times i {\hat e}_2 + a_i^{m,k} \times k {\hat e}_3 \ .
\label{threedimpoly}
\fqr
The base reduced entries of $a_i^{m,k}$ may be attributed into 'colors' or a group
theory indices labeling a representation.

Next the volume $V$ and the $\partial V$ surface area of the polytope
region is deduced from the entries $a_i^{m,k}$.  The volume is the
sum of the individual entries $a_i^{m,k}$ over the entire lattice,

\bqr
p_j = a_{i,(j)} 2^i  \quad V_s= \sum_{i,k,m} a_i^{k,m} \ .
\fqr
The surface area of the polytope is a region bounded by the entries
of the entries $a_i^{m,k}$.  The bounded region is found via the
differences of the entries $a_i$; in two dimensions,

\bqr
V_{sf} = \sum_{ij} \vert a_i^{j} - a_{i-1}^{j}\vert -
  \sum_{ij} \vert a_i^{j} - a_i^{j-1} \vert \ .
\fqr
The region bounding the polytope is deduced from the differences
in the integers.
 
The terms in both series, $V_s$ and $V_{sf}$, are defined or
computed via the expansions,

\bqr
P_1^i = \sum M_{(1)}^{ij} p^j \qquad
P_2^i = M_{(2)}^{ij} p^j = \sum \vert a_i-a_{i-1}\vert_{{\rm p int}}
\ ,
\fqr
\bqr
P_1^i= \sum a_i\vert_{\rm p int} \ ,
\fqr
defined for the integer p configuration.
Even though the the individual terms $\vert a_i-a_i\vert$
in the summations involved the expansion are absolute value, the
entire sum is found via a summation over the individual numbers
$p$ parameterizing the lattice and its configuration.  (A discussion 
of the computation of the matrices $M$ is found in \cite{ChalmersOne}.)

The 'colored' boundary is given a boundary via the same formalism,
but with a generalized difference $\vert a_i-a_{i-1}\vert$;
group theory or 'color' differences found with a different inner product
are possible.  The summations for these numbers may also be inverted
to obtain the values $a_i$ in terms of $p^j$ and an associated matrix.

An example list of this variables is given in the following table,

\bqr
\pmatrix{ p & a_i & p_1 & p_2 \cr 1 & 1 & 1 & 1 \cr
 2 & 01 & 1 & 2 \cr
 3 & 11 & 2 & 0 \cr
 4 & 001 & 1 & 2 \cr
 5 & 101 & 2 & 2 \cr
 6 & 011 & 2 & 2 \cr
 7 & 111 & 3 & 0 \cr } \ .
\label{partitions}
\fqr
The number $p$ is listed, followed by the binary format; the integers
$p_1$ and $p_2$ are the sums $\sum a_i$ and $\sum \vert a_i - a_{i+1}\vert$,
in a cyclic fashion around the numbers $p$.

The polyhedron is constructed by the single numbers spanning the
multiple layers in 3-d, or by one number with the former grouped as
$p_1 p_2 \ldots p_n$.  The generalization to multiple dimensions is
straightforward.  Also, the sewing and disection of the manifolds 
based on the operations on the integer $p$ is straightforward.

\vskip .2in 
\noindent {\it Conclusions}

Manifolds are put in one to one correspondence with the integers via 
an embedding into a d-dimensional lattice.  To each integer $z$ there 
is a set of latticized points specifying the manifold.  

Properties such as cohomology, and other topological ones, are formulated 
in a uniform sense via mappings 

\bqr  
P(\tilde z) = \sum b_i {\tilde z}^i \qquad \tilde z = \sum c_i z^i \ . 
\fqr 
The first function specifies the index of the manifold $z$ via its value on $z$.  
The latter function is a redefinition of the coordinates labeling the lattice 
configurations; these two functions should enter with eachother for each of 
the topological indices.   
 
The values of the indices, specified on all manifolds within a lattice of size 
$N^d$, may be put into correspondence with an L-series via, 
 
\bqr 
\zeta(s,C) = \prod \left(1+a_p p^{-s} + p^{1-2s}\right)^{-1} \ ,   
\label{ellipticzeta}
\fqr 
and $-p+b_p=a_p$; a simple ordering is for the primes to be put into correspondence 
with the $z$-values and $b_p=P(z)$.   The function ${\tilde z}=f(z)$ is chosen to 
systematize the entries in \rf{ellipticzeta}.  In this fashion, there is an elliptic 
curve that specifies the topological index in $d$-dimensions for a lattice of size 
$V$ (and as $V\rightarrow \infty$).

\vfill\break

\end{document}